\documentclass[twocolumn,10pt,prl,nofootinbib]{revtex4-1}
\usepackage[utf8]{inputenc}

\usepackage{hyperref}
\usepackage{soul}
\usepackage{array}

\usepackage[dvips]{graphicx}
\usepackage{epsfig,amsmath,amssymb,verbatim,mathrsfs,array,layout,textcomp,amssymb,latexsym,slashed,graphicx,booktabs,color,mathtools,tikz}
\usepackage{cleveref}
\usepackage{tikz-feynman}

\newcommand{\beq}{\begin{eqnarray}}
\newcommand{\eeq}{\end{eqnarray}}

\def\beqa{\begin{eqnarray}}
\def\eeqa{\end{eqnarray}}

\newcommand{\ev}{\end{array}\right)}
\newcommand{\bmtwo}{\left(\begin{array}{cc}}
\newcommand{\bmthree}{\left(\begin{array}{ccc}}
\newcommand{\emn}{\end{array}\right)}
\newcommand{\bmtwoc}{\left\{\begin{array}{cc}}
\newcommand{\bmthreec}{\left\{\begin{array}{ccc}}
\newcommand{\emnc}{\end{array}\right\}}
\newcommand{\ba}{\begin{array}}
\newcommand{\ea}{\end{array}}

\newcommand{\noblettersym}{N}
\newcommand{\alklettersym}{s}

\newcommand{\nobletter}{\textrm{\noblettersym}}
\newcommand{\alkletter}{\textrm{\alklettersym}}

\newcommand{\Bvec}{{\bf B}}
\newcommand{\Snobvec}{{\bf{\noblettersym}}}
\newcommand{\Salkvec}{{\bf{\alklettersym}}}
\newcommand{\gammaalk}{{\gamma_\alkletter}}
\newcommand{\gammanob}{{\gamma_\nobletter}}
\newcommand{\Mnob}{B_{\nobletter}}
\newcommand{\Bz}{B}
\newcommand{\Bx}{\delta B_x}
\newcommand{\By}{\delta B_y}

\newcommand{\Bcomp}{B_c}
\newcommand{\BRoW}{B_R}

\newcommand{\Malk}{B_{\alkletter}}
\newcommand{\balkvec}{{\bf b}_{\alkletter} }
\newcommand{\bnobvec}{{\bf b}_{\nobletter} }
\newcommand{\Rpu}{R_{\rm pu}}
\newcommand{\spu}{{\bf s}_{\rm pu}}
\newcommand{\Ralk}{\Gamma_\alkletter}
\newcommand{\Rnob}{\Gamma_\nobletter}
\newcommand{\Rnobtensor}{\mathcal{R}_\nobletter}
\newcommand{\Ralktensor}{\mathcal{R}_\alkletter}

\newcommand{\Rnobz}{\Gamma_{\nobletter,z}}
\newcommand{\Ralkz}{\Gamma_{\alkletter,z}}
\newcommand{\Rnobalk}{R_{\nobletter\alkletter}}
\newcommand{\Ralknob}{R_{\alkletter\nobletter}}

\newcommand{\Szalk}{\alklettersym_z}
\newcommand{\Sznob}{\noblettersym_z}
\newcommand{\Spalk}{\alklettersym_+}
\newcommand{\Spnob}{\noblettersym_+}

\newcommand{\Sperpnob}{{\bf \noblettersym}_\perp}
\newcommand{\Sxalk}{\alklettersym_x}
\newcommand{\Sxnob}{\noblettersym_x}
\newcommand{\Syalk}{\alklettersym_y}
\newcommand{\Synob}{\noblettersym_y}
\newcommand{\bnobx}{b_{\nobletter,x}}
\newcommand{\bnoby}{b_{\nobletter,y}}

\newcommand{\bnobp}{b_{\nobletter,+}}
\newcommand{\bnobperp}{{\bf b}_{\nobletter,\perp}}
\newcommand{\Bp}{\delta B_{+}}
\newcommand{\Bperp}{\delta{\bf B}_{\perp}}

\newcommand{\oalk}{\omega_\alkletter}
\newcommand{\onob}{\omega_\nobletter}
\newcommand{\oalknot}{\omega_{0\alkletter}}
\newcommand{\onobnot}{\omega_{0\nobletter}}

\newcommand{\SxalkR}{\alklettersym_{x}'}

\newcommand{\Lnew}{l_{\rm R}}
\newcommand{\oRoW}{\Omega_{\rm R}}
\newcommand{\Lvec}{\mathbf{l}}
\newcommand{\Lz}{l}

\def\lsim{\mathrel{\rlap{\lower4pt\hbox{\hskip1pt$\sim$}}
 \raise1pt\hbox{$<$}}} 
\def\gsim{\mathrel{\rlap{\lower4pt\hbox{\hskip1pt$\sim$}}
 \raise1pt\hbox{$>$}}}

\begin{document}
\font\mini=cmr10 at 0.8pt

\title{A Rotating-Wave Comagnetometer Detector for Particle Physics}
\author{Itay M. Bloch}
\affiliation{Physics Division, Lawrence Berkeley National Laboratory, Berkeley, CA 94720, USA}
\affiliation{Berkeley Center for Theoretical Physics, Department of Physics, University of California, Berkeley, CA 94720, USA}
\author{Or Katz}
\affiliation{School of Applied and Engineering Physics, Cornell University, Ithaca, New York 14853, USA}

\begin{abstract}
Many extensions of the Standard Model propose the existence of new particles or forces, aiming to answer mysteries such as the identity of the elusive dark matter. Atomic-based detectors are at the forefront of technologies designed to search for these particles or forces through their couplings to fermions, enabling the testing of well-motivated models, such as axion-like particles, which could form dark matter. These detectors also probe new long-range interactions between the detectors and spin-polarized objects, as well as interactions mediated by light particles that break CP symmetry, introducing a coupling between the detector and an unpolarized object. However, the sensitivity of these detectors is often constrained by magnetic noise, limiting their effectiveness to a narrow region of parameter space. We propose and develop a technique, which we name the Rotating Wave comagnetometer (RoW comag), that can suppress magnetic noise at tunable frequencies while maintaining high sensitivity to target signals, significantly expanding the potential reach of these detectors. We analyze its operation for testing various extensions to the Standard Model and show how it could improve current sensitivities by several orders of magnitude. This work paves the way for a new class of tabletop experiments aimed at searching for new physics, including the exploration of well-motivated axion-like particle dark matter models at higher masses than previously attainable.
\end{abstract}

\maketitle
\section{Introduction}
The Standard Model serves as the foundational framework for our understanding of the universe, yet it is known to be incomplete. One of the most significant gaps in our understanding is the existence of Dark Matter (DM)~\cite{P5report}. Based on gravitational evidence observed across various astronomical scales, DM is believed to constitute the majority of the matter energy-density in the universe~\cite{PDG}. However, the particle nature of DM—including its mass and non-gravitational interactions—remains unknown.

A well-motivated class of DM models associates DM with Axion-Like Particles (ALPs)~\cite{Abbott:1982af,Preskill:1982cy,Dine:1982ah}. These particles extend the concept of QCD-Axions~\cite{Peccei1977,Weinberg1978,Wilczek1978} and appear in many theories beyond the Standard Model. In some scenarios, ALPs may explain the observed asymmetry between matter and antimatter~\cite{Co:2019jts,Co:2020xlh}. ALPs could interact non-gravitationally with the spins of fermions such as neutrons, protons, or electrons. If ALPs constitute DM, they are expected to induce an oscillatory energy splitting of spins aligned with their velocity vector at a frequency proportional to their mass, potentially allowing for their detection through their coupling to matter. QCD-Axions, ALPs, and many other models of new light particles also suggest a similar energy shift of spins in the presence of nearby spin-polarized~\cite{Moody:1984ba,Liao:2007ic,Dobrescu:2006au,VasilakisThesis,Arkani-Hamed:2004gbh} or even unpolarized materials~\cite{Moody:1984ba,Arvanitaki:2014dfa,OHare:2020wah,Pospelov:1997uv}. 

\begin{figure*}[t]\includegraphics[width=1.0\textwidth]{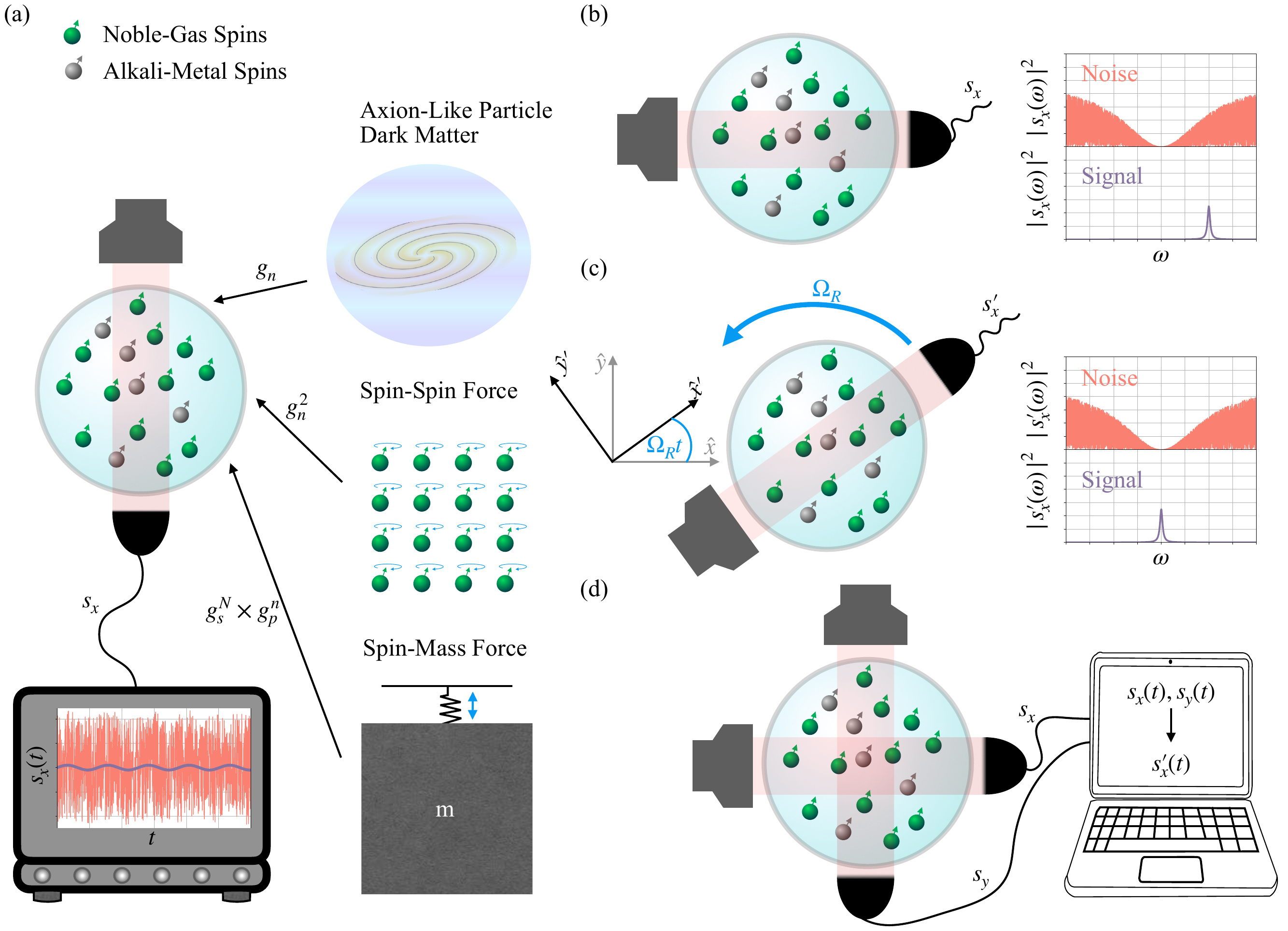}
\caption{\textbf{Comagnetometer particle physics detectors.} (a) Spin-based detector. A glass cell containing spin-polarized alkali-metal and noble-gas atoms is used to detect new particles or forces that couple to the noble-gas spins. We analyze three scenarios: Axion-Like Particle Dark Matter, long-range spin-spin interactions, and spin-mass couplings, with the generated signal proportional to $g_n$, $g_n^2$ and $g_s^N\times g_p^n$ respectively. In these configurations, an optical beam is used to detect small signals associated with these interactions. The detector's performance is constrained by its noise level, often magnetic field noise. (b) Self-compensating comagnetometers are among the most sensitive detectors at low frequencies, where magnetic noise is highly suppressed. However, their performance deteriorates rapidly for signal frequencies above the sub-hertz range. (c) By leveraging the concept of non-inertial frame transformations through rotation at a frequency $\Omega_{R}$, the Rotating-Wave comagnetometer (RoW comag) suppresses magnetic noise at nonzero frequencies with appropriately tuned control fields, enabling precise detection of small oscillatory signals (see text). (d) A dual-probe measurement in the laboratory frame eliminates the need for physical rotation, facilitating efficient estimation of measurements in the non-inertial frame. }
\label{fig:compdetector}
\end{figure*}

n recent years, small-scale terrestrial experiments have shown great capabilities in searching for new physics\cite{Bloch:2019lcy,Bloch:2021vnn,Bloch:2022kjm,Gavilan-Martin:2024nlo,Wu:2019exd,Garcon:2019inh,Wei:2023rzs,Xu:2023vfn,Jiang:2021dby,Abel:2017rtm,Abel:2022vfg,Lee:2022vvb,QUAX:2024fut,Ouellet:2018beu,Gramolin:2020ict,Devlin:2021fpq,Thomson:2019aht,Grenet:2021vbb,Friel:2024shg,Sulai:2023zqw,Arza:2021ekq,Fedderke:2021aqo,Zhang:2023qmu}. However, much of the parameter space probed by these experiments has already been excluded by astrophysical observations. For example, the coupling of ultralight ALPs to nucleons across a broad range of masses is strongly constrained by measurements of supernova explosions~\cite{Zyla:2020zbs,Chang:2018rso,Carenza:2019pxu} or the cooling of neutron stars~\cite{Beznogov:2018fda,Hamaguchi_2018,Keller_2013,Sedrakian_2016,Buschmann:2021juv}. Among the most sensitive technologies for ALP DM detection are sensors based on atomic ensembles of spins~\cite{Bloch:2019lcy,Bloch:2021vnn,Bloch:2022kjm,Gavilan-Martin:2024nlo,Wu:2019exd,Garcon:2019inh,Wei:2023rzs,Xu:2023vfn,Jiang:2021dby,Abel:2017rtm,Abel:2022vfg,Lee:2022vvb,Zhang:2023qmu}. Self-compensating comagnetometers~\cite{kornack2002dynamics,KornackThesis,Kornack:2004cs,Vasilakis:2008yn,VasilakisThesis,Brown:2010dt,BrownThesis} stand out among the most sensitive spin-based detectors, currently providing the strongest limits on ALP-neutron couplings for masses between $10^{-16}$ and $2\times 10^{-14}~{\rm eV}/c^2$~\cite{Bloch:2019lcy,Wei:2023rzs,Lee:2022vvb,Gavilan-Martin:2024nlo}. These detectors have also been employed in searches for violations of Lorentz symmetry~\cite{Brown:2010dt,Smiciklas:2011xq}, new long-range spin-spin forces~\cite{Vasilakis:2008yn,Almasi:2018cob}, and novel spin-mass forces~\cite{Lee:2018vaq}, among other phenomena~\cite{Afach:2021pfd,GNOME:2023rpz}.

The sensitivity and suitability of the self-compensating comagnetometer as a particle detector largely stem from its inherent ability to suppress background magnetic noise~\cite{kornack2002dynamics}. However, the bandwidth of this suppression is extremely narrow and centered near zero frequency, restricting effective noise suppression to sub-millihertz frequencies. As a result, the detector is primarily effective for probing small ALP masses. This detector currently offers the highest sensitivity for ALP-DM at masses $\lesssim 2\times 10^{-14}~{\rm eV}/c^2$. However, at higher masses (and correspondingly, higher frequencies), its magnetic subtraction becomes insufficient, and non-compensated resonant comagnetometers offer better sensitivity. Successfully enabling subtraction at those frequencies would facilitate effective searches in previously unexplored regions of the parameter space.

We propose and develop a technique that extends the operational frequency range of comagnetometers, thereby broadening the range of ultra-light DM masses that can be detected. This technique, which we call the Rotating-Wave comagnetometer (RoW comag), significantly expands the frequencies over which the detector can effectively suppress background fields while maintaining sensitivity to signals associated with new particles. We evaluate the potential sensitivity of this approach for searching for new physics and demonstrate that it opens new avenues for exploring ultralight ALP DM, and long-range spin-spin or spin-mass forces, enhancing detection capabilities by several orders of magnitude.

\subsection{Self-Compensating Comagnetometers}

\textit{Detection Capability.} Self-compensating comagnetometers~\cite{kornack2002dynamics} use two overlapping atomic ensembles contained within a glass cell, as shown in Fig.~\ref{fig:compdetector}a. The first ensemble consists of an alkali-metal vapor with controllable and optically readable electron spins, functioning as a high-precision optical magnetometer~\cite{kornack2002dynamics,KornackThesis,VasilakisThesis,BrownThesis}. The second ensemble is composed of dense noble gas atoms with nonzero nuclear spins, characterized by coherence times ranging from minutes to hours~\cite{Bloch:2019lcy,VasilakisThesis,KornackThesis,BrownThesis}. The two ensembles are spin-polarized along a static magnetic field $\Bz\hat{z}$, through optical pumping and spin-exchange collisions~\cite{Walker:1997zzc}. The interaction of new-physics forces or fields with the fermions in the ensembles induces collective spin flips, which manifest as a tilting of the ensembles' collective spin vectors. Tilts of the alkali-metal ensemble are detectable as a magnetometer signal through optical readout.

In Fig.~\ref{fig:compdetector}b, we illustrate the configuration for detecting the coupling of ultralight ALP DM with the noble gas nuclear spins (variants for other Particle Physics models are presented in the Methods section). With the ALP's de-Broglie wavelength much longer than the detector dimensions, the gradient of the ALP field can couple to the nuclear spins as a uniform (anomalous) magnetic field $\bnobvec(t)$~\cite{OldWind}. $\bnobvec(t)$ oscillates at a frequency equal to the ALP kinetic energy $\hbar \omega_{\textrm{DM}}\approx m_{\rm DM}c^2$~\cite{OldWind,Bloch:2019lcy}. The nuclear collective-spin vector $\Snobvec$, initially polarized along the $z$-direction, is rotated by components of this field in the $xy$ plane ($\bnobperp=\bnobx\hat{x}+\bnoby\hat{y}$) thus exciting the spins off axis and generating nonzero transverse spin components ($\Sperpnob=\Sxnob\hat{x}+\Synob\hat{y}$). These components, in turn, apply a magnetic torque on the alkali-metal collective spin vector $\Salkvec$, initially polarized along the $z$-direction as well, via the spin-exchange (Fermi-contact) interaction~\cite{Walker:1997zzc}, generating nonzero spin components in the $xy$ plane. Therefore, the alkali-metal spins become susceptible to the ALP DM field $\bnobperp$, and satisfy the linear relationship $|\Sxalk|=|\chi_{b}\bnoby|$ at low frequencies, with the susceptibility $\chi_{b}$ given in Eq.~\eqref{eq:anomsus} in the Methods. The polarized alkali-metal ensemble, being a circular birefringent optical medium for light tuned near the dipole transition~\cite{WalkerProbe}, rotates the polarization vector of an optical probe beam passing through the cell across $\hat{x}$. Polarimetry of the probe beam, whose signal $S_x$ is proportional to the alkali-metal spin component, ${s}_x$, enables detection.

\textit{Noise Suppression Mechanism.} The detector sensitivity is limited by background noise, typically magnetic fields $\Bperp$ oscillating at $\omega_{\textrm{DM}}$ in the $xy$ plane, which would rotate the noble-gas or alkali-metal spins off axis, leading to false detection. For example, in a magnetically-shielded environment at finite temperatures, the electrically conductive shields produce magnetic field noise at levels of $|\Bperp|\sim10\,\textrm{fT}/\sqrt{\textrm{Hz}}$~\cite{Lee:2007be}. Self-compensating comagnetometers suppress such fields at low frequencies by leveraging the fact that, unlike ALP DM fields, the magnetic noise couples simultaneously to both the noble gas and the alkali-metal spins at strengths proportional to their gyromagnetic ratios $\gamma_{\nobletter}$ and $\gamma_{\alkletter}$, respectively. 

Accounting for the hybrid alkali-metal and noble gas dynamics, we can express the total alkali-metal spin response to magnetic field noise in the frequency domain as
\begin{equation}\label{eq:sx_response}
s_x(\omega)=\tfrac{1}{2}\left(\chi_B(\omega)\delta B_+(\omega)+\chi_B^*(-\omega)\delta B^*_+(-\omega)\right).
\end{equation}
Using complex notation, $\Bp=\Bx+i\By$ represents the magnetic field noise vector that rotates counter-clockwise in the $xy$ plane, and $\delta B_+^*(-\omega)\equiv(\Bp(-\omega))^*$\ is the orthogonal field vector that rotates clockwise. The complex response function $\chi_B(\omega)$ is approximately given by 
\begin{equation}\label{eq:chi_B}
\chi_B(\omega)=\frac{\Delta_\nobletter+\nu}{-J^2+\Delta_\nobletter(i\Ralk+\Delta_\alkletter)}\gammaalk \Szalk,
\end{equation}
where $\Gamma_{\alkletter}$ denotes the alkali-metal spin decoherence rate, $J$ is the collective spin-exchange rate~\cite{Shaham:2021qvu,Katz:2022jxc} and $\Delta_\nobletter(\omega,B)=\omega_{0\nobletter}+\gamma_{\nobletter} B-\omega$ and $\Delta_{\alkletter}(\omega,B)=\omega_{0\alkletter}+l+\gamma_{\alkletter} B-\omega$ are the relative detunings of the field oscillating at frequency $\omega$ from the noble-gas and alkali-metal magnetic (Larmor) frequencies, respectively. $l$ represents light induced shift by optical beams. $\omega_{0\nobletter},\,\omega_{0\alkletter}$ represent the spin-exchange shifts applied by one ensemble on the other and $\nu\equiv\omega_{0\alkletter}\gammanob/\gammaalk$.

The response to transverse magnetic field noise is significantly suppressed at low frequencies when a static magnetic field $\Bcomp\hat{z}$ satisfying $\Delta_{\nobletter}(\Bcomp)=-\nu$ is applied, resulting in $\chi_B(\omega=0)=\chi_B^*(-\omega=0)=0$ in Eq.~\eqref{eq:chi_B}, as shown in Fig.~\ref{fig:compdetector}b. This suppression arises from adiabatic dynamics of the dual ensemble, where the noble-gas collective spin vector adiabatically follows its total field in the $xy$ plane. The total field felt by each ensemble is composed of the magnetic noise, anomalous fields and the spin-exchange field exerted by the other ensemble. At $B=\Bcomp$, the spin-exchange field generated by the noble gas nuclei countereacts the magnetic field noise with equal magnitude, leading to maximal suppression of the magnetic noise detected by the alkali-metal spins, and consequently, by the detector.

\begin{figure}[ht!]
\includegraphics[width=0.5\textwidth]{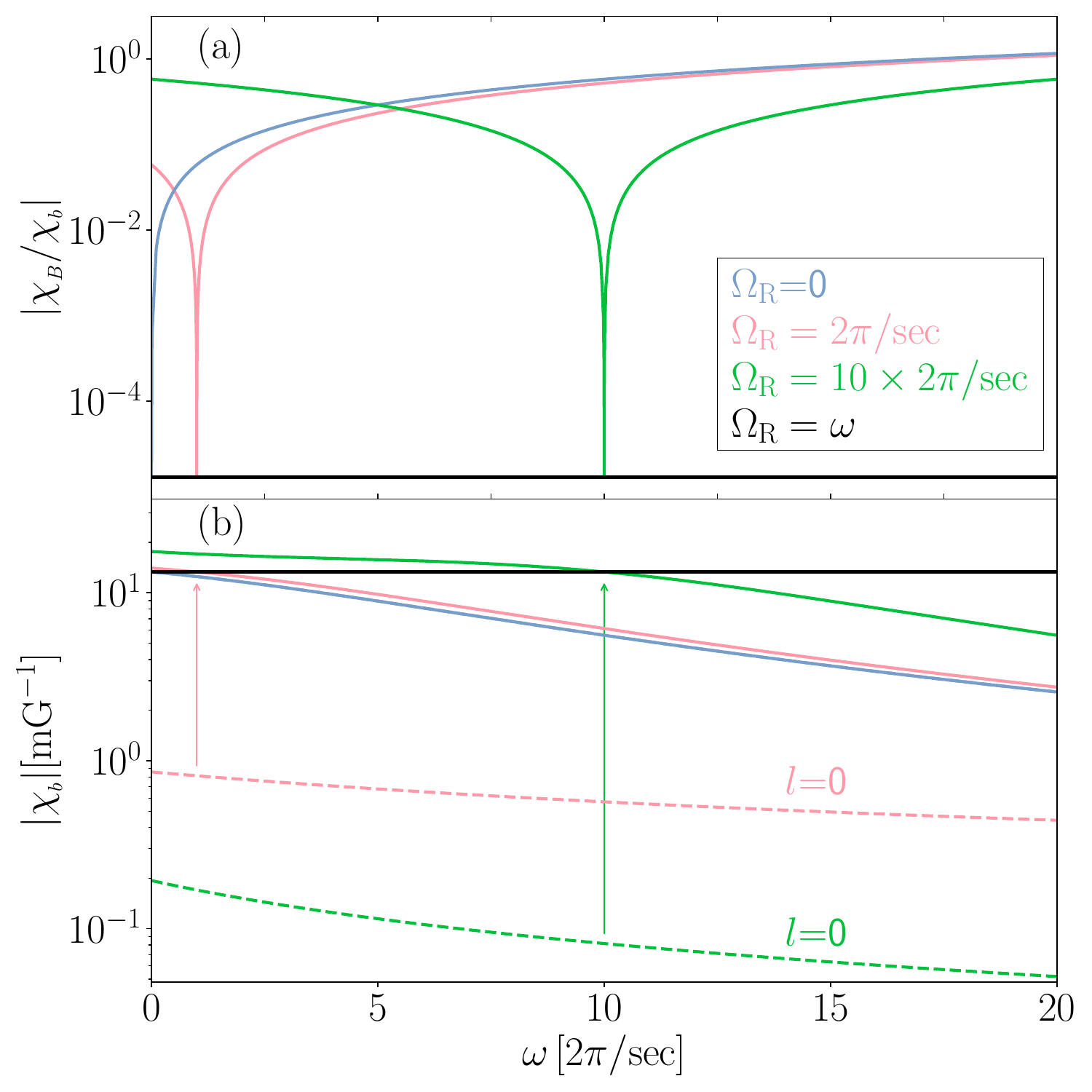}
\caption{\textbf{Response to magnetic noise and signal}. (a) The response to magnetic noise, characterized by the susceptibility $\chi_B$, shows strong suppression at low frequencies for the non-rotating detector (blue) and at higher frequencies determined by the rotation frequency (red, green). The static magnetic field is tuned to $B=B_R$ (see Eq.~\eqref{eq:BR}) which only matches the field of self-compensating comagnetometers when $\Omega_R=0$. The suppression of magnetic noise is shown relative to the response function $\chi_b$ for an anomalous signal field, representing interactions with new particles or forces (see text and methods). The maximal suppression is constrained by decoherence and collisional processes (see Eq.\eqref{eq:magsus}). (b) The response to the signal field can achieve a high, $\oRoW$-independent value when $\omega=\Omega_R$ through optimal tuning of the light-shift $l$ (see Eq.~\eqref{eq:LR}). This tuning increases the dashed lines (zero light-shift) to solid lines ($l=l_R$). Black lines in (a) and (b) illustrate the performance for $\Omega_R=\omega$ with the application of optimal control fields $B=B_R$ and $l=l_R$.}
\label{fig:newperformance}
\end{figure}

\begin{figure*}[p!]
\includegraphics[width=0.9\textwidth]{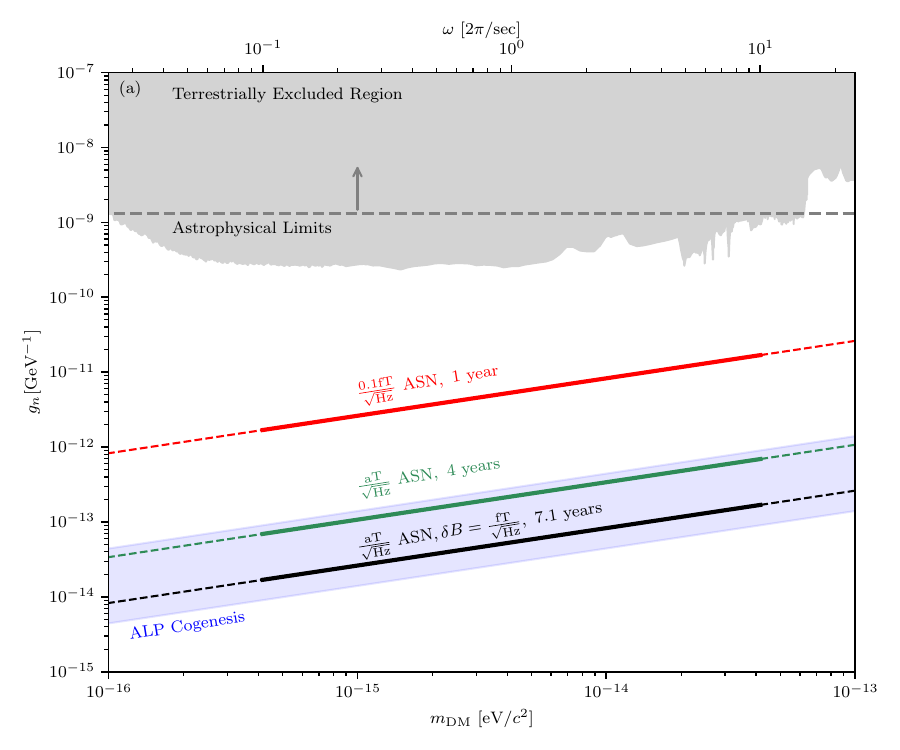}
\includegraphics[width=0.45\textwidth]{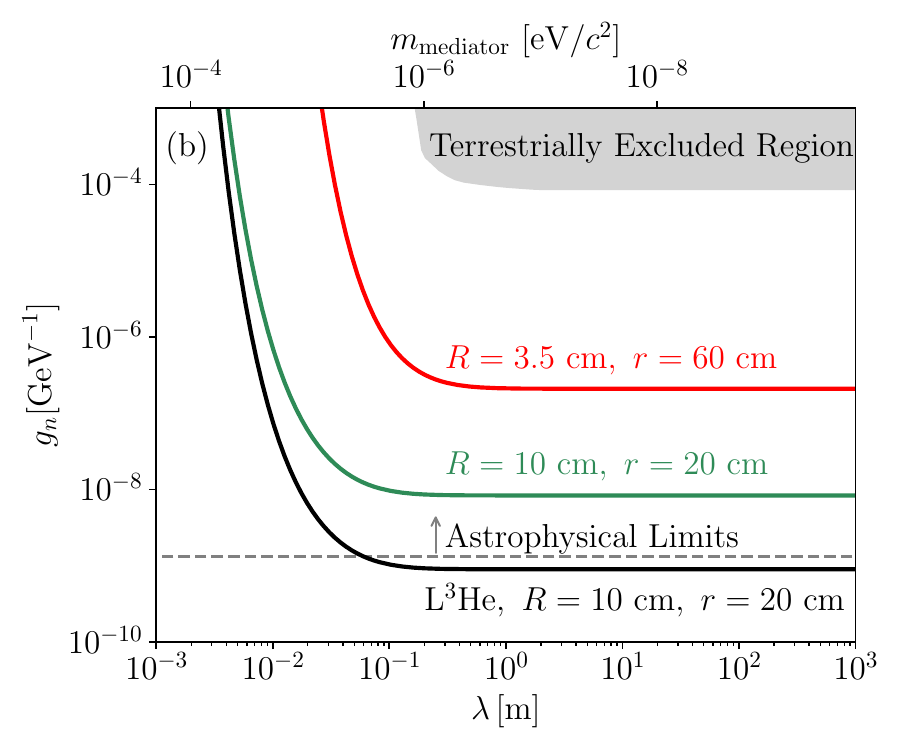}
\includegraphics[width=0.45\textwidth]{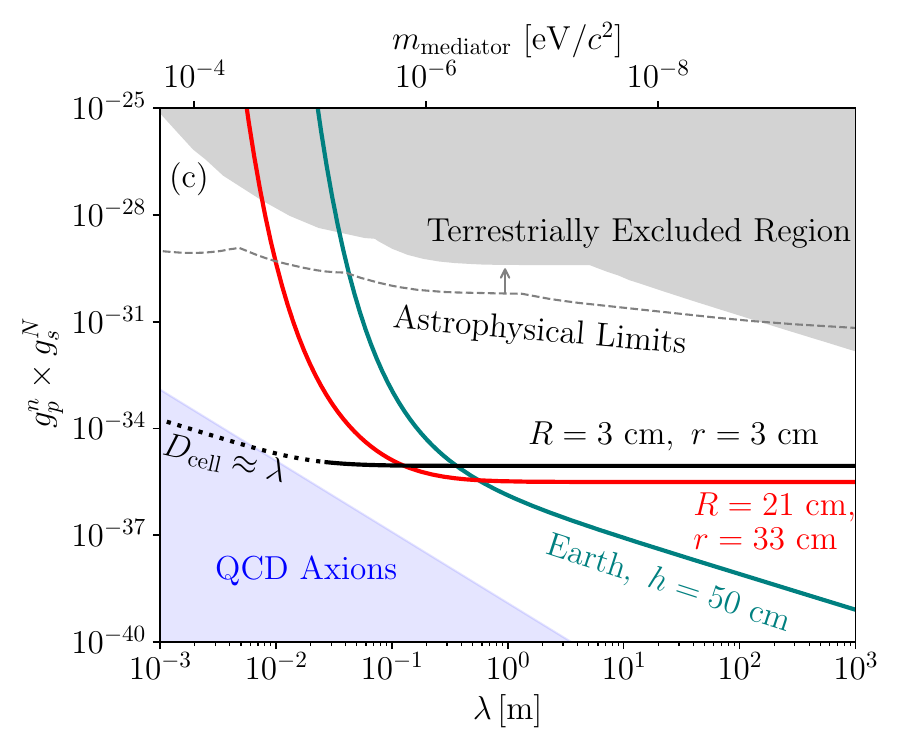}
\caption{Projected sensitivity and constraints for Beyond-Standard-Model Interactions. (a) Axion-Like Particle (ALP) Dark Matter (DM) coupling to neutrons, (b) long-range spin-spin forces, and (c) spin-mass coupling. Projections for different feasible configurations, as detailed in Table~\ref{tab:setups}, are shown in red, teal, and black lines. Grey-shaded regions represent excluded parameters based on terrestrial detectors~\cite{Bloch:2019lcy,Zhang:2023qmu,Tullney:2013wqa,Lee:2018vaq,Vasilakis:2008yn,VasilakisThesis,Xu:2023vfn,Wei:2023rzs}, while dashed grey lines denote constraints from astrophysical observations~\cite{Raffelt:2012sp,Buschmann:2021juv,OHare:2020wah}. The light blue region highlights the theoretically motivated parameter space from ALP-Cogenesis models~\cite{Co:2019jts,Co:2020xlh} in (a) and QCD axion models with CP-violating couplings~\cite{Arvanitaki:2014dfa} in (c). $m_{\rm{DM}}$ denotes the ALP mass in (a) and $\lambda$ denotes the range of coupling for which the interaction decays exponentially in (b) and (c), see Eqs.~\eqref{eq:spin_spin_bn} and~\eqref{eq:spin_mass_bn} respectively. $\omega=m_{\rm DM}c^2/\hbar$ is the corresponding frequency of the ALP-DM field in (a), and $m_{\rm mediator}=\hbar /(\lambda c)$ is the mass of the interactions' mediator in (b) and (c).}
\label{fig:sensitivity}
\end{figure*}

In practice, the maximal suppression factor $\chi_B^{-1}$ is finite, and is determined by the noble-gas decoherence and collisional pumping processes; see Eq.~\eqref{eq:magsus} in the Methods section. For Ref.~\cite{VasilakisThesis}'s experimental configuration, the maximal suppression is about five orders of magnitude with a bandwidth of about $\omega\lesssim 200\,(2\pi)\mu$Hz, as illustrated in Fig.~\ref{fig:newperformance}a (blue line). For higher frequencies, Eq.~\eqref{eq:chi_B} accurately describes the magnetic noise suppression and its degradation at higher noise frequencies, with barely any suppression at $\omega$ of a few hertz~\cite{PhysRevLett.95.230801}. This result originates primarily from the inability of the collective noble-gas spin vector to adiabatically follow its time-varying total field and generate a spin-exchange field that cancels the alkali-metal response to the noise.

Notably, controlling external parameters such as $B$ or $l$ does not enable tuning of the noise suppression to higher frequencies. This limitation arises because, for any nonzero frequency, the condition $\Delta_{\nobletter}(B)=\nu$ cannot be satisfied simultaneously for both the $+\omega$ and $-\omega$ components in Eq.~\eqref{eq:sx_response}, leaving at least one of the rotating noise field vectors unsuppressed. 

\section{Rotating-Wave Comagnetometer}
\subsection{Self compensation of magnetic noise}
Suppressing magnetic noise at finite frequencies would seem to require the cancellation of both the clockwise and counter-clockwise rotating noise components, which have response amplitudes $\chi_B(\omega)$ and $\chi_B^*(-\omega)$, respectively. However, for stationary fields at $\omega=0$, both terms are cancelled simultaneously. This observation motivates us to consider the Self-compensating comagnetometer equations of motion in a rotating frame of reference that rotates counterclockwise at frequency $\oRoW$. By analyzing this non-inertial frame, we propose a straightforward scheme that achieves noise suppression without requiring any physical rotation. The coordinates in the rotating frame are given by
\begin{eqnarray}
 \hat{x}'	=\cos(\Omega_{\text{R}}t)\hat{x}+\sin(\Omega_{\text{R}}t)\hat{y},\label{eq:x_prime}\\
\hat{y}'=\cos(\Omega_{\text{R}}t)\hat{y}-\sin(\Omega_{\text{R}}t)\hat{x}.
\end{eqnarray} In this frame of reference, any fields oscillating counterclockwise at frequency $\omega=\oRoW$ would appear stationary, as illustrated in Fig.~\ref{fig:compdetector}c. The equations of motion in the rotating frame include a non-inertial ''rotation field" $-\Omega_{\textrm{R}}\hat{z}$, which couples to both the noble-gas and alkali-metal spins. This coupling results in a uniform shift in their resonance frequencies $\Delta_{\nobletter}$ and $\Delta_{\alkletter}$ by $-\oRoW$. We express the $\hat{x}'$ spin in the rotating frame, $\SxalkR$, as a function of the \textit{laboratory} noise fields \begin{align}
 \label{eq:sx_prime_response}
s_{x}^{\prime}(\omega)&=\tfrac{1}{2}\bigl( \chi_B(\omega+\Omega_{\textrm{R}})\delta B_+(\omega+\Omega_{\textrm{R}}) \\+&\chi_B^{*}(\Omega_{\textrm{R}}-\omega)\delta B^{*}_+(\Omega_{\textrm{R}}-\omega)\bigr ),\notag
\end{align}
where crucially, both the clockwise and counter-clockwise rotation frequencies are shifted equally by the same negative rate $\Omega_{\textrm{R}}$. Consequently, the stationary spin signal in the rotating frame, $s_x'(\omega=0)$, is susceptible to laboratory magnetic fields oscillating at \textit{nonzero} frequency $\Omega_{R}$ in the laboratory frame, with the same susceptibility magnitude $|\chi_B(\Omega_{\textrm{R}})|$. Tuning the static magnetic field to
\begin{equation}\label{eq:BR}
B_{\textrm{R}}=\Bcomp-\Omega_{\textrm{R}}/\gamma_{\nobletter},
\end{equation}
enables correction for the rotation fields, ensuring maximal suppression $\chi_B(\Omega_{\textrm{R}})\rightarrow 0$ to compensate for noisy fields oscillating at the target frequency $\Omega_{\textrm{R}}$, unlike the self-compensating comagnetometers which only suppress nearly static magnetic field noise. The suppression of this rotating frame magnetometer for nonzero rotation frequencies is presented in Fig.~\ref{fig:newperformance}a.

Operating the detector at a magnetic field $B_{\textrm{R}}$ affects the magnetometer's response to both noisy and anomalous fields. While the change in $\Delta_{\nobletter}$ for the noble-gas spins is compensated by adjusting $B$, it shifts the resonance frequency of the alkali-metal spins $\Delta_{\alkletter}(B)$. Nevertheless, the application of a light-shift term 
\begin{equation}\label{eq:LR}
 l_R=-(\gamma_{\alkletter}/\gamma_{\nobletter}-1)\Omega_{\textrm{R}},
\end{equation} compensates for the alkali-metal frequency shift without affecting the noble-gas spins, thereby restoring the magnetometer's sensitivity to its maximal value. This renders the response to ALP fields at $\omega=\oRoW$ independent of $\Omega_{\textrm{R}}$, as shown in Fig.~\ref{fig:newperformance}b. We therefore conclude that a RoW comag maintains its sensitivity to ALP fields, enabling effective suppression of magnetic field noise at the rotation frequency, given optimal magnetic field and light-shift control.

Notably, the rotating frame signal can be reconstructed in the laboratory frame without any physical rotation. Instead, simultaneous measurement of $s_x(t)$ and $s_y(t)$ is required. The rotating frame signal, $s_x'$, can then be reconstructed using Eq.~\eqref{eq:x_prime} through either digital post-processing of measurements or by passing the analog signals through lock-in amplifiers and combining their in-phase and out-of-phase components, as illustrated in Fig.~\ref{fig:compdetector}d. This technique provides an efficient method for practically measuring the rotating frame signal in the laboratory frame.

\subsection{Performance Estimation}
We estimate the projected performance of the detector in the detection of ultralight ALP DM coupled to neutrons (Fig.~\ref{fig:sensitivity}a), long-range ALP-mediated spin-spin forces (Fig.~\ref{fig:sensitivity}b), and long-range ALP-mediated spin-mass forces (Fig.~\ref{fig:sensitivity}c). The exact details assumed in the derivation of these constraints (including the conversion from coupling constants to anomalous magnetic signals) are given in the Methods and in particular in Table~\ref{tab:setups}. Here, we briefly explain the main assumptions made. In each plot, we present the projected performance of the RoW comagnetometer with solid red, teal, and black lines under different assumptions. The dashed lines of the same colors could potentially be probed by the detector, though exploring them is beyond the scope of this work. Existing astrophysical bounds~\cite{Raffelt:2012sp,Buschmann:2021juv,OHare:2020wah} are presented in dashed gray, and the currently excluded region by terrestrial experiments~\cite{Bloch:2019lcy,Zhang:2023qmu,Tullney:2013wqa,Lee:2018vaq,Vasilakis:2008yn,VasilakisThesis,Xu:2023vfn,Wei:2023rzs} is the shaded light gray region. For Figs~\ref{fig:sensitivity}a and~\ref{fig:sensitivity}b, we also present predictions of particularly motivated theoretical models~\cite{Co:2019jts,Co:2020xlh,Arvanitaki:2014dfa} in a shaded blue region.

When searching for ALP DM, the detector searches for an oscillating signal at an unknown frequency, sourced by the DM field. To achieve this, $\oRoW$ is scanned via a set of measurements with $\oRoW$s equally spaced between $0.1\times 2\pi~{\rm Hz}$ and $10\times{2\pi}~{\rm Hz}$. In practice, this requires adaptation of the control fields in Eq.~\eqref{eq:BR} and Eq.~\eqref{eq:LR} at each measurement point, and the dual probe measurement which enables reconstructing $\SxalkR$ for any individual $\oRoW$. The combined projected constraint is shown in Fig.~\ref{fig:sensitivity}a. The bottom horizontal axis indicates the mass of the DM particle, and the vertical axis the coupling of ALPs to neutrons, denoted by $g_{n}$. The anomalous field is proportional to $g_n$ (see Eq.~\eqref{eq:bn_ALP}). We take each measurement at a given $\oRoW$ to last the full (red, black) or part of (teal) the ALP coherence time. For measurements reaching the full coherence time, elongating the measurement time $t$ would scale the projected bound according to $t^{-1/4}$, while for shorter measurement the improvement scales as $1/\sqrt{t}$. The time indicated in Fig.~\ref{fig:sensitivity} is the total time of all different points.

In Fig.~\ref{fig:sensitivity}b,c we present the projected detector sensitivity to long-range spin-spin force and spin-mass forces, respectively, with $\lambda\equiv \hbar /(m_{\rm mediator}c)$ representing the range of interaction. We focus the analysis on ALP-mediated forces, though one could easily recast the same search's sensitivity in the context of other SM extensions with similar phenomenology~\cite{Moody:1984ba,Liao:2007ic,Dobrescu:2006au,VasilakisThesis,Arkani-Hamed:2004gbh}. In these cases, all measurements are assumed to take 3 years, and the detector is assumed to be magnetically shielded and placed near a sphere filled with spin-polarized or unpolarized nuclei.  The nuclei or their spins assumed to move or precess at a frequency $\oRoW$ that is tuned to minimize the detector's technical noise. For the spin-polarized source the anomalous field is proportional to $g_n^2$ while for the unpolarized source it scales as $g_s^N\times g_p^n$ as given by Eq.~\eqref{eq:spin_spin_bn}, Eq.~\eqref{eq:spin_mass_bn}, respectively. Here, $g_{p}^n=2m_n c^2 g_n$, while $g_s^N$ is the dimensionless CP-violating scalar coupling of the ALP to nucleons. Here, measuring for a longer time $t$ would indefinitely improve $g_n^2$ and $g_s^N\times g_p^n$ as $\sqrt{t}$.

\section{Discussion}

In this work we presented a novel method relying on the effective (or physical) rotation of a comagnetometer to drastically reduce magnetic noise at a controlled frequency. Other comagnetometry systems have been placed on rotating stages in the past~\cite{Zhang:2023qmu,Brown:2010dt}. However, in those cases, a static signal was modulated, and the need for physical rotation required adding mechanical stages to the basic comagnetometer setup. We demonstrated how by modifying the detector parameter, and adding a second probe beam, a search for oscillating signals is facilitated, which could be of great benefit in many particle physics searches.

The presented detector can potentially improve current sensitivities by four orders of magnitude through suppression of magnetic field noise which is a major limiting factor. Reaching this performance would require optimization of the optical magnetometer readout, which is influenced by factors such as the finite number of photons in the polarimeter, the degree of circular birefringence in the medium, and technical electronic noises. However, the only major change in the detector setup which is necessary for demonstrating the new method is the addition of a second probe beam. With this simple modification, the RoW comag should be able to explore models which no terrestrial or astrophysical probe is sensitive to. Additional feasible improvements would allow the detector to probe several highly motivated extensions to the Standard Model~\cite{Co:2019jts,Co:2020xlh,Arvanitaki:2014dfa}. We focused on three extensions to the Standard Model: ALP-DM coupled to neutrons, ALP-mediated spin-spin long-range interactions, and ALP-mediated spin-mass long-range interactions. However, other configurations of the RoW comag, or generalizations of it in other systems could potentially probe other extensions to the SM as well.

The true fundamental quantum noise of comagnetometers is associated with the finite number of noble-gas atoms, and is exceptionally low (on the order of $0.01~\rm{aT}/{\sqrt{\rm Hz}}$). Reducing practical noises is of key importance in achieving that sensitivity. Therefore, this work represents a significant milestone towards a system that could utilize the full potential of atomic spin detectors.

\clearpage

 \onecolumngrid
\part*{\centerline{Materials and Methods}}

\section*{Time dependent dynamics of the Rotating Wave comagnetometer detector}

The comagnetometer consists of two overlapping spin ensembles: an alkali-metal vapor with a mean spin vector $\Salkvec$, and a noble-gas ensemble with a mean spin vector $\Snobvec$. Their dynamics, described in Ref.~\cite{kornack2002dynamics}, evolve according to the Bloch equations:
\begin{eqnarray}
\dot{\Salkvec}&=&\gammaalk\left(\Bvec+\Mnob\Snobvec+\balkvec+\Lvec/\gammaalk \right) \times\Salkvec
+\left(\Rpu \spu+ \Ralknob\Snobvec-\Ralktensor\Salkvec\right),
\\
\dot{\Snobvec}&=&\gammanob (\Bvec+\Malk\Salkvec+\bnobvec) \times \Snobvec+ \Rnobalk {\Salkvec}-\Rnobtensor \Snobvec.
\end{eqnarray}

We focus on a configuration where the applied magnetic field, pump beam, and applied light-shift are aligned along the $\hat{z}$-axis, similar to other comagnetometers. In the limit of small transverse fields, the steady-state axial spin components are given by:
\begin{equation}\label{eq:szalk}
\Szalk=\frac{\Rpu\Rnobz}{2(\Rnobz\Ralkz- \Rnobalk\Ralknob)},
\end{equation} 
and 
\begin{equation}\label{eq:sznob}
\Sznob=\Szalk\Rnobalk/\Rnobz.
\end{equation}

Assuming these axial components remain constant during the dynamics allows us to simplify the problem and focus on the evolution of the transverse spin components. For conciseness, we use the complex notation 
$\Spalk=\Sxalk+i\Syalk$, and $\Spnob=\Sxnob+i\Synob$ for the transverse spin components, and
$\Bp=\Bx+i\By$, $\bnobp=\bnobx+i\bnoby$ for the transverse noise and signal fields. With this notation, the dynamics can be expressed as:
\begin{eqnarray}
\partial_t\Spalk&=&(i\oalk-\Ralk)\Spalk+(\Ralknob-i\gammaalk\Mnob\Szalk)\Spnob-i\gammaalk\Szalk\Bp\\
\partial_t\Spnob&=&
(i\onob-\Rnob)\Spnob+(\Rnobalk-i\gammanob\Malk\Sznob)\Spalk
-i\gammanob\Sznob\left(\Bp+\bnobp\right),
\end{eqnarray}
where we defined
\begin{eqnarray}
\oalk&=&\gammaalk\left(\Bz+\Mnob\Sznob\right)+\Lz=\gammaalk\Bz+\oalknot,\\
\onob&=&\gammanob\left(\Bz+\Malk\Szalk\right)=\gammanob\Bz+\onobnot.
\end{eqnarray}

Owing to the spectrally narrow nature of ALPs, applying a Fourier transform to these equations allows us to focus on the response to these frequencies, which is relevant for signal detection. This transformation eliminates terms related to initial conditions and can be expressed as:
\begin{equation}\label{eq:mateqs}
\begin{pmatrix}
i(\omega-\oalk)+\Ralk & i\gammaalk\Mnob\Szalk-\Ralknob \\
i\gammanob\Malk\Sznob-\Rnobalk & i(\omega-\onob)+\Rnob
\end{pmatrix}
\begin{pmatrix}
\Spalk (\omega) \\
\Spnob (\omega)
\end{pmatrix}
=
\begin{pmatrix}
-i\gammaalk\Szalk\Bp(\omega) \\
-i\gammanob\Sznob \Bp(\omega)-i\Sznob\gammanob\bnobp(\omega).
\end{pmatrix}
\end{equation}

Solving this system by matrix inversion results in the response of the spins to the anomalous and noise fields. Since only the alkali-metal spins are measured, we focus on their response (which incorporates the coupling to noble gas spins),
\begin{equation}\label{eq:ssteady}
\Spalk(\omega)=(\chi_B(\omega)\Bp(\omega)+\chi_b(\omega)\bnobp(\omega)).
\end{equation}
The linear response function to anomalous fields coupled directly to the noble-gas spins is given by 
\begin{equation}\label{eq:anomsus}
\chi_b(\omega)=-\frac{\left(\gammaalk\Mnob\Szalk+i\Ralknob\right)}{\left(i(\omega-\oalk)+\Ralk\right)\left(i(\omega-\onob)+\Rnob\right)-\left(i\gammanob\Malk\Sznob-\Rnobalk\right)\left(i\gammaalk\Mnob\Szalk-\Ralknob\right)}\gammanob\Sznob,\end{equation}
and the linear response function to magnetic field noise is given by
\begin{equation}\label{eq:magsus}
\chi_B(\omega)=\frac{(\omega-\onob-\gammanob\Mnob\Sznob)-i(\Rnob+\gammanob\Sznob\Ralknob/(\gammaalk\Szalk))}{\left(i(\omega-\oalk)+\Ralk\right)\left(i(\omega-\onob)+\Rnob\right)-\left(i\gammanob\Malk\Sznob-\Rnobalk\right)\left(i\gammaalk\Mnob\Szalk-\Ralknob\right)}\gammaalk\Szalk.\end{equation}
Therefore, the expression for $\chi_B$ indicates the sensitivity of the comagnetometer to magnetic noise, while $\chi_b$ reflects its sensitivity to anomalous signals. These parameters define the response characteristics of the detector. The expression for $\chi_B$ given in the main text is the result when neglecting the noble-gas related relaxation mechanisms ($\Rnob,\Ralknob,\Rnobalk$), which practically set a lower limit on the detector bandwidth and maximal magnetic field suppression.

\section{Optimal control fields}
Maximizing the sensitivity of the detector to anomalous fields requires optimizing the ratio $|\chi_b(\omega)/\chi_B(\omega)|$. From Eq.~\eqref{eq:anomsus} and Eq.~\eqref{eq:magsus}, this ratio is maximized for $\omega=\oRoW$ when $\onob=\oRoW-\gammanob\Mnob\Sznob$. This corresponds to a magnetic field $\Bz=\BRoW$ with 
\begin{equation}
\BRoW\equiv-\Mnob\Sznob-\Malk\Szalk+\frac{\oRoW}{\gammanob}=\Bcomp+\frac{\oRoW}{\gammanob}.
\end{equation}
In a frame rotating at $\oRoW$, one reproduces the picture of the standard self-compensating comagnetometer of Ref.~\cite{kornack2002dynamics}, where static fields are suppressed. Signals at $\oRoW$ shift to zero, while the noble gas perceives an effective shift equivalent to a change in the magnetic field amplitude by $\oRoW/\gammanob$ relative to the compensation point. Adjusting the external field by this amount reproduces the compensation-point behavior, with the alkali-metal ensemble experiencing a modified field. This adjustment ensures an optimal ratio between the sensitivity to signal and magnetic noise at $\oRoW$.

While tuning $\BRoW$ effectively minimizes sensitivity to ambient magnetic fields, other noise sources (such as electronic or probe polarization noise) become more dominant when the amplitude of $\chi_b(\omega)$ is decreased. To ensure that $\chi_b(\omega)$ is not decreased by the change of magnetic field to $\Bz=\BRoW$, we optimize it as a function of $\Lvec=\Lz\hat{z}$, yielding the optimal value 
\begin{equation}
\Lnew(\oRoW)=-\left(\frac{\gammaalk}{\gammanob}-1\right)\oRoW+\frac{\left(\Mnob\Rnobalk+\Malk\Rnob\right)\left(\gammaalk\Rnob\Szalk+\gammanob\Ralknob\Sznob\right)}{\left(\Rnob^2+\gammanob^2\Mnob^2\Sznob^2\right)}\approx -\left(\frac{\gammaalk}{\gammanob}-1\right)\oRoW.
\end{equation}

For the parameters we consider, $\Rnob,\Ralknob,\Rnobalk$ are small and their effect on the signal response is negligible, leading to the approximate response
\begin{equation}
\Spalk(\oRoW)\approx-\frac{\Szalk\gammaalk\bnobp}{(i\Ralk+(\Lz-\Lnew(\oRoW))}.
\end{equation}
Achieving maximal sensitivity thus requires tuning $\Lz$ with a resolution of approximately $\Ralk$, which corresponds to the alkali-magnetometer bandwidth. As the light-shift is proportional to $\oRoW$, achieving this for large $\oRoW$ would require high-power and far-detuned lasers. However, as many noise sources (including the atomic shot noise and magnetic noise) are suppressed in the same manner when $\Lz=0$, in some cases tuning $\Lz=\Lnew$ may not be crucial for optimal performance.

\section{Sensitivity to Transient Signals
}
We briefly address the detector's response to transient signals, which may arise from sudden changes in system conditions. While the steady-state solution in Eq.~\eqref{eq:ssteady} describes the long-term behavior, transient noise can be seen as producing exponentially decaying homogeneous solutions. Efficiently vetoing periods of transient noise in a practical search is crucial, and longer veto durations reduce the overall measurement time. A shorter coherence time for the system's modes allows for quicker recovery from transient events.

At the optimal control parameters, coupling between the two spin ensembles leads to large decoherence rates for both hybrid modes of the spins. These rates can be approximated as $\frac{\gammaalk \Malk \Szalk}{\gammaalk \Malk \Szalk + \gammanob \Mnob \Sznob}\Ralk, \frac{\gammanob \Mnob \Sznob}{\gammaalk \Malk \Szalk + \gammanob \Mnob \Sznob}\Ralk$. These are identical to the decoherence rates in the self-compensating comagnetometers of Ref.~\cite{kornack2002dynamics}. This indicates that both spin ensembles have coherence times primarily determined by the alkali-metal ensemble, enabling efficient noise vetoing and rapid system recovery. If light-shift compensation is not applied, at higher $\oRoW$s, one of the modes becomes longer lived, providing another motivation for applying light-shift compensation.

\section*{Anomalous magnetic-like fields}\label{sec:signal}
Modifications to the Standard Model can introduce anomalous fields, which can manifest as either irreducible backgrounds or as signals under specific conditions. Here, we examine three such models: ALPs as DM, long-range spin-spin interactions, and spin-mass couplings. Throughout this section, we use natural units where $\hbar=1$ and $c=1$.
\subsection{Axion-Like Particles as Dark Matter}
In the model we consider in this work, ALP DM couples to neutron spins through a coupling constant $g_{n}$. The coherence time of the ALP-induced field is inversely proportional to its energy uncertainty, approximately $\sim 2\pi \cdot 1.6\times 10^6/m_{\rm DM}$, allowing it to be treated as a monochromatic field over short measurement periods. The anomalous field generated by the ALP is given by~\cite{OldWind,Bloch:2019lcy}
\begin{equation}\label{eq:bn_ALP}
\bnobvec=\epsilon_{nN} g_{n}\sqrt{{2\rho_{\rm DM}}}{\bf v}_{\rm DM} \cos(\langle E_{\rm DM}\rangle t+\theta_0)/\gammanob,
\end{equation}
with ${\bf v}_{\rm DM}$ denoting the DM velocity relative to the lab-frame, $\rho_{\rm DM}$ denoting the local DM energy density, and $\theta_0$ is an unknown initial phase. We take $|{\bf v}_{\rm DM}|=238{\rm km}/{\rm sec}$, and $\rho_{\rm DM}=\sqrt{0.4{\rm GeV}/{\rm cm}^3}$. For practical purposes, the ALP energy $\langle E_{\rm DM}\rangle$ is approximated by $m_{\rm DM}$. The parameter $\epsilon_{nN}$ represents the fraction of the noble gas spin due to neutron-spin contributions, which is approximately 100\% for $^{3}\textrm{He}$~\cite{Stadnik:2014xja}.
This effective coupling arises from an interaction described by the Lagrangian:
\begin{equation}
\mathcal{L}=g_{n}\partial_\mu a\bar{n}\gamma^{\mu}\gamma^5n,
\end{equation}
where $a$ represents the ALP field, $
n$ is the neutron Dirac spinor, and $\gamma^{\mu}$ are the Dirac gamma matrices. 

\subsection{Long-Range Spin-Spin Interactions}
New long-range spin-spin interactions are predicted by a variety of SM extensions~\cite{Moody:1984ba,Liao:2007ic,Dobrescu:2006au,VasilakisThesis,Arkani-Hamed:2004gbh}. In particular, ALPs, whether they are DM or not, introduce such interactions. We focus on ALP-induced forces, though our proposed experimental setup could be reanalyzed in the context of other extensions as well. We consider a spin-polarized source based on nuclear spins with large neutron fraction ${\bf I}_{\rm source}$ located at distance ${\bf r}$ from the RoW comag. The induced anomalous field coupled to the noble gas nuclei in the detector is given by~\cite{VasilakisThesis} \begin{equation}
\bnobvec\approx \frac{\epsilon_{Nn}g_{\rm ann}^2}{2\pi\gammanob}\left(\left(\frac{1}{r^2\lambda}+\frac{1}{r^3}\right){\bf I}_{\rm source}-(\hat r\cdot {\bf I}_{\rm source})\hat{r}\left(\frac{1}{\lambda^2r}+\frac{3}{r^2\lambda}+\frac{3}{r^3}\right)\right)e^{r/\lambda},
\end{equation}
where $\lambda=1/m_{\rm mediator}$ is the interaction range, with ${m}_{\rm mediator}$ as the ALP mediator's mass. When $\lambda\ll r$, the interaction decays exponentially, while for $\lambda\gg r$, it approximates a dipole-dipole interaction, scaling as $1/r^3$. We assume for simplicity that the spin source is a sphere of radius $R$ with a uniform spin density $n_{\rm source}$, and a uniform spin-polarization $P_{\rm source}$ (taken to be 100\%). We take the contribution of neutron spins to the spin of the nuclei as $\epsilon_{n,{\rm source}}$ . Because at low frequencies, various noise sources are considerably larger, we assume that the spin source precesses under a magnetic field at a frequency $\oRoW$ such that its polarization along the $\hat{x}$ axis is modulated by $\cos(\oRoW t)$. While all three components of ${\bf I}_{\rm source}$ contribute to the anomalous field in all directions, for simplicity we focus on the contribution from $I_{{\rm source},x}$ to the anomalous field. Considering this component only, and $\oRoW\ll 1/r,m_{\rm mediator}$, the induced anomalous field is given by
\begin{equation}\label{eq:spin_spin_bn}
\bnobx=\frac{1}{\gammanob}g_{ann}^2 n_{\rm source}\epsilon_{n,{\rm source}}\epsilon_{nN}\frac{(R+\lambda)e^{-(r+R)/\lambda}+(R-\lambda)e^{-(r-R)/\lambda}}{2r^3}\left((r+\lambda)^2+\lambda^2\right)\cos(\oRoW t).
\end{equation}
For simplicity, in our estimates we take $\cos(\oRoW t)\hat{x}\to \hat{x}'(t)$ (see Eq.~\eqref{eq:x_prime}).

\subsection{Spin-Mass Couplings}
ALPs or QCD axions with CP-violating couplings can produce spin-mass interactions. Assuming a single atom of mass number A at distance $r$ from the detector, the induced field is~\cite{Arvanitaki:2014dfa}
\begin{equation}
\bnobvec= \frac{Ag_{s}^N g_p^n}{8\pi m_n\gammanob}\epsilon_{nN}\left(\frac{1}{r\lambda}+\frac{1}{r^2}\right)e^{-r/\lambda}\hat{r},
\end{equation}
where $\lambda=1/m_{\rm mediator}$, $m_n$ is the mass of a neutron, $g_p^n=2m_ng_{n}$ is the dimensionless (pseudoscalar) coupling of ALPs to neutron spins, and $g_{s}^N$ denotes the (CP-violating) scalar coupling to nucleons in the massive object (assumed to be equal for protons and neutrons). For $r\gg \lambda$, the interaction decays exponentially, while for $r\ll \lambda$, it decays like $1/r^2$.

We assume for simplicity the mass is always placed in the $\hat{x}'(t)$ (see Eq.~\eqref{eq:x_prime}) direction. Similar sensitivity is achieved if the distance (rather than relative direction) between the mass and RoW comag is modulated. We further take the mass to be a uniform sphere of radius $R$ and density $\rho_{\rm source}$, and so it generates a field at the detector's center for $r\gtrsim R$
\begin{equation}\label{eq:spin_mass_bn}
{\bf b}_N= \frac{g_{s}^N g_p^n\rho_{\rm source}}{4 m_n^2\gammanob r^2}\epsilon_{nN}\lambda(r+\lambda)\left((R+\lambda)e^{-(r+R)/\lambda}+(R-\lambda)e^{-(r-R)/\lambda}\right)\hat{x}'(t).
\end{equation}
This estimate assumes a point-like detector and $\oRoW\ll 1/r,m_{\rm mediator}$. 

\section*{Proposed Setups and Projected performance}\label{sec:setups}

In this section, we detail the proposed configurations that we considered in the calculation of the projected performance shown in Fig.~\ref{fig:sensitivity}. Each plot describes the sensitivity to one of the three models we consider, and presents three possible realizations of the sensor. We begin by discussing general assumptions and parameters, followed by details of the specific setups, with further information shown later in Table~\ref{tab:setups}.

The comagnetometer parameters, such as the decoherence rates and magnetization, are based on the detector in Ref.~\cite{VasilakisThesis}, except where otherwise specified. We consider two levels of white magnetic noise within the relevant parts of the spectrum: (1) $|\delta B|\approx10~{\rm fT}/{\sqrt{Hz}}$ achievable using high-permeability $\mu$-metal shielding limited primarily by Johnson noise~\cite{Lee:2007be}, and (2) $|\delta B|\approx1~{\rm fT}/{\sqrt{\rm Hz}}$, achievable with low-noise shields such as ferrites or superconducting shields, the latter limited by thermal radiation~\cite{kornack2007low}. Ferrites are likely to provide such sensitivity only for higher frequencies~\cite{kornack2007low}.

Another source of noise stems from the limited number of alkali-metal atoms and the variance in their collective spin, known as the Atomic Shot Noise (ASN). Neglecting the effects of $\Rnob,\Rnobalk$ and $\Ralknob$, the ASN is given by~\cite{VasilakisThesis} $\delta\Sxalk=\sqrt{{\xi}/{(2\Ralk n_\alklettersym V_{\rm eff}})},$ where $\xi$ is a dimensionless parameter which depends on $\Szalk$, and equal to $0.65$ for $\Szalk=0.25$. $n_{\alkletter}$ is the number density of the alkali-metal ensemble, and $V_{\rm eff}$ is the effective probe beam volume in the cell. For the RoW comganetometer detector operating at $\Lnew$ and $\BRoW$, our ASN remains consistent with that of the non-rotating system in Ref.~\cite{VasilakisThesis}. In units of magnetic field, this noise can be expressed as 
\begin{equation}\label{eq:ASN}
\delta B_{\rm ASN,eff}=\frac{1}{\gammaalk \Szalk}\sqrt{\frac{\xi\Ralk}{2 n_\alkletter V_{\rm eff}}}.
\end{equation}

Photon shot noise, associated with the finite number of photons, can be tuned to be smaller than ASN~\cite{PhysRevLett.95.063004}, making ASN the limiting noise source. We examine two regimes: one with a mild $10\%$ improvement over Ref.~\cite{VasilakisThesis} ($|\delta B_{\rm ASN,eff}|\approx0.1~{\rm fT}/{\sqrt{\rm Hz}}$) and another with improved sensitivity $|\delta B_{\rm ASN,eff}|\approx1~{\rm aT}/{\sqrt{\rm Hz}}$ which could be achieved using larger cells and higher temperatures~\cite{RomalisaT}.

We now discuss the parameters for Fig~\ref{fig:sensitivity}. For the long-range interaction searches, the oscillation frequency is known and chosen to minimize technical noise. However, when searching for ALP DM we scan over $\oRoW$ to search for the signal, because the mass of ALP-DM is unknown. Therefore, we consider the bandwidth of our sensor for a fixed $\BRoW,\Lnew$, as well as the range of frequencies we expect to measure, determined by practical considerations of $\Lnew$.

Since the maximal magnetic subtraction is below the lowest quantum noise, the dependence of $\chi_B$ on $\Rnob,\Rnobalk,\Ralknob$ plays no role. Therefore, we can estimate the square root of the power spectral density of the alkali-metal spin $\Spalk$ at the vicinity of $\omega=\oRoW$ as 
\begin{equation}
\delta \Spalk(\omega=\oRoW+\delta \omega)\approx\sqrt{\frac{\xi}{2\Ralk n_\alklettersym V_{\rm eff}}+\left(\frac{\delta \omega\gammaalk\Szalk}{\Ralk\gammanob\Mnob\Sznob}\delta B(\omega)\right)^2},
\end{equation}
with the first term associated with ASN, and the second term with imperfect suppression of the magnetic noise $\delta B$ at a frequency $\Omega_R+\delta\omega$. Assuming a frequency independent $\delta B(\omega)$, we consider the effective bandwidth, BW, as the range of $\delta\omega$s for which the second term is smaller than the first,
\begin{equation}
{\rm BW}=2\cdot \left(\frac{\delta B_{\rm ASN,eff}}{{\rm aT}/{\sqrt{\rm Hz}}}\right)\left(\frac{\delta B}{{\rm fT}/{\sqrt{\rm Hz}}}\right)^{-1}\left(\frac{\gammanob\Mnob\Sznob}{2\pi \times 17 {\rm Hz}}\right)\cdot 2\pi \times 17~{\rm mHz}.
\end{equation}
Therefore, lower magnetic background noise $\delta B$, would effectively increase the bandwidth around a given $\oRoW$. A broader measurement bandwidth would allow to fully scan a predetermined region of frequencies $[\omega_{\rm min},\omega_{\rm max}]$ with fewer points, thus decreasing the required total measurement time needed for a fixed targeted sensitivity. 

The projected constraints shown in Fig~\ref{fig:sensitivity}a are constructed via consideration of multiple independent measurement points $i=1,...,N$, each assumed to be sensitive to frequencies $(\oRoW)_i\pm {\rm BW}/2$ with the $(\oRoW)_i$s equally spaced by ${\rm BW}$ within the scan range. Each of these measurements is performed for a duration of $2\pi p/((\oRoW)_i$-{\rm BW}/2), where the number of periods for each frequency, $p$, can be inferred from the total measurement time we took. We keep $p\leq 1.6\times 10^{6}$, as going beyond that only marginally improves the sensitivity for longer measurement times, as explained in the main text. The specific parameters taken in each projected search are provided in Table~\ref{tab:setups}.

\renewcommand{\arraystretch}{1.5}

\begin{table}[t]
\centering
\begin{tabular}{|c|c|c|c|c|c|c|}
\hline
\raisebox{1.5ex}{(1a) line} & \raisebox{1.5ex}{${\delta B}_{\rm ASN,eff}~\left[\tfrac{\rm aT}{\sqrt{\rm Hz}}\right]$} & \raisebox{1.5ex}{${\delta B}~\left[\tfrac{\rm fT}{\sqrt{\rm Hz}}\right]$} & \raisebox{1.5ex}{$t~\left[{\rm years}\right]$} & \raisebox{1.5ex}{Periods ($p$)} & \raisebox{1.5ex}{${\rm BW}~\left[\tfrac{2\pi}{\rm sec}\right]$} & \raisebox{1.5ex}{$\oRoW~\left[\tfrac{2\pi}{\rm sec}\right]$}\rule{0pt}{5ex} \\
\hline\hline
\textcolor{red}{\bf Red line} & 100 & 10 & 1 & $1.6\times 10^6$ & 0.34 & 0.1-10 \\
\hline
\textcolor{teal}{\bf Teal line} & 1 & 10 & 4 & $9.5 \times 10^4$ & 0.0034 & 0.1-10 \\
\hline
{\bf Black line} & 1 & 1 & 7.1 & $1.6\times 10^6$ & 0.034 & 0.1-10 \\
\hline
\noalign{\vskip 15pt}
\hline
\raisebox{1.5ex}{(1b) line} & \raisebox{1.5ex}{${\delta B}_{\rm ASN,eff}~\left[\tfrac{\rm aT}{\sqrt{\rm Hz}}\right]$} & \raisebox{1.5ex}{${\delta B}~\left[\tfrac{\rm fT}{\sqrt{\rm Hz}}\right]$} & \raisebox{1.5ex}{$t~\left[{\rm years}\right]$} & \raisebox{1.5ex}{$n_{\rm source}P_{\rm source}~\left[{\rm cm}^{-3}\right]$} & \raisebox{1.5ex}{$R~\left[{\rm cm}\right]$} & \raisebox{1.5ex}{$r-R~\left[{\rm cm}\right]$}\rule{0pt}{5ex} \\
\hline\hline
\textcolor{red}{\bf Red line} & 1 & 10 & 3 & $3\times 10^{20}$ & 3.5 & 56.5 \\
\hline
\textcolor{teal}{\bf Teal line} & 1 & 10 & 3 & $3\times 10^{20}$ & 10 & 10 \\
\hline
{\bf Black line} & 1 & 10 & 3 & $2.6\times 10^{22}$ & 10 & 10 \\
\hline
\noalign{\vskip 15pt}
\hline
\raisebox{1.5ex}{(1c) line} & \raisebox{1.5ex}{${\delta B}_{\rm ASN,eff}~\left[\tfrac{\rm aT}{\sqrt{\rm Hz}}\right]$} & \raisebox{1.5ex}{${\delta B}~\left[\tfrac{\rm fT}{\sqrt{\rm Hz}}\right]$} & \raisebox{1.5ex}{$t~\left[{\rm years}\right]$} & \raisebox{1.5ex}{$\rho_{\rm source}~\left[\tfrac{\rm gram}{{\rm cm}^{3}}\right]$} & \raisebox{1.5ex}{$R~\left[{\rm cm}\right]$} & \raisebox{1.5ex}{$r-R~\left[{\rm cm}\right]$}\rule{0pt}{5ex} \\
\hline\hline 
\textcolor{red}{\bf Red line} & 1 & 10 & 3 & 11.3 & 21 & 12 \\
\hline
\textcolor{teal}{\bf Teal line} & 1 & 10 & 3 & 2.5 & $6.4\times 10^{8}$ & 50 \\
\hline
{\bf Black line} & 1 & 10 & 3 & 11.3 & 3 & 0 \\
\hline
\end{tabular}
\caption{The parameters assumed for each of the nine setups presented in Fig.~\ref{fig:sensitivity}. We assume throughout that $\epsilon_{nN}=\epsilon_{\rm n,{\rm source}}=1$, and that other parameters match those of Ref.~\cite{VasilakisThesis} unless otherwise specified. See text for further discussion.}\label{tab:setups}
\end{table}

The red line of Fig.~\ref{fig:newperformance}b is based on the long-range spin-spin interaction search for Ref.~\cite{VasilakisThesis}, with improvements associated with the lower noise that can be realized at higher frequencies, higher alkali-metal densities, and assumed unity $^3$He spin-source polarization. The black solid line shows the sensitivity for a fully polarized liquid $^{3}$He sample.

The red line of Fig.~\ref{fig:newperformance}c assumes a similar unpolarized source to the SMILE experiment (which had two 250~kg Pb bricks)~\cite{Lee:2018vaq}, with the improvement achieved due to the lower noise that can be realized at higher frequencies, and the higher alkali-metal densities. The teal line uses the earth as the mass source, with the detector oscillating with an effective distance of $50~{\rm cm}$. We remark that unlike all other configurations discussed in this work, this setup requires the physical oscillation of the detector to modulate the signal. Finally, the black line assumes a Pb sphere, oscillating such that its edge reaches a negligible distance to the detector (much smaller than $\lambda$). The dotted black line is when $\lambda$ becomes smaller than the typical cells used, and therefore, one needs to either account for the fact that a gradient is formed, or use smaller cells, where the ASN will be bigger. Interestingly, within that region, a CP-violating interaction of the QCD axion may be probed~\cite{Arvanitaki:2014dfa}. This perhaps motivates focusing on this setup, and resolving its issues, in order to probe this highly motivated part of the parameter space.

\section{Acknowledgements}

This work was performed in part at Aspen Center for Physics, which is supported by National Science Foundation grant PHY-2210452. This work was performed in part at the Aspen Center for Physics, which is supported by a grant from the Simons Foundation (1161654, Troyer).

\bibliography{bibliography}

\end{document}